\documentstyle[12pt]{article}

\begin{document}
\centerline{\bf  Electrodynamics in the presence of an axion}

\vskip 1in
\centerline {C. Corian\`{o} *}

\centerline {Institute for Theoretical Physics}
\centerline{and}
\centerline{Department of Physics}
\centerline { State University of New York at Stony Brook}
\centerline {Stony Brook, NY 11794  USA}

\vspace{2cm}
\centerline{Mod. Phys. Lett, to appear}
\newcommand{\newc}{\newcommand}
\newc{\beq}{\begin{equation}}
\newc{\eeq}{\end{equation}}
\newc{\beqa}{\begin{eqnarray}}
\newc{\eeqa}{\end{eqnarray}}
\newc{\grad}{\nabla}
\newc{\Tc}{T_{c}}
\newc{\dmu}{\partial_{\mu}}
\newc{\dnu}{\partial_{\nu}}
\newc{\half}{\frac{1}{2}}
\newc{\quarter}{\frac{1}{4}}
\newc{\parm}{\par\medskip}
\newc{\ul}{\underline}
\newc{\del}{\partial}
\vskip 0.5cm
\begin{abstract}
The low energy limit of an axion field coupled to gauge fields
is investigated through the behaviour of the gauge fields propagator
in a local vacuum angle background.
The local (singular) part of the effective action for the axion field is
calculated at one loop level.
In the case of a timelike,
linearly growing axion field,
representing a massive axion, we give an asymptotic expansion
of the causal propagator and we solve nonlocally for the first
coefficient. We show that, for a generic axionic background,
short distance propagation of the gauge field
is well defined.
\end{abstract}
\vskip 3cm
\centerline{Supported by the NSF under grant NSF 91 08054}
\centerline{ and by the Department of Energy under Grant No.
\ DE-FG02-88ER40388 }

*A. Della Riccia fellow
\newpage

\section{Introduction}

One of the most interesting aspects of instanton effects in QCD is related
to the strong CP problem and why the vacuum angle is extremely small
\cite{t'Hooft}.

Various solution have been proposed in the literature for this problem.
In one scheme one can set $\theta =0 $ at tree level and then assume that
higher order effects in the magnitude of $\theta$ are negligible
\cite{Di Vecchia}, while
a second option is to introduce a massless quark so that $\theta$
becomes unobservable. This second possibility is ruled out
phenomenologically.

A third alternative is to introduce a dynamical variable $\theta (x)$
in such a way that an effective interaction
\beq
L_{int}=c\theta (x)F\tilde{F}
\eeq
(c being a constant)

dynamically sets the
 vacuum angle $\theta_{QCD}$
to be zero \cite{Peccei}.

This third approach, generally known as the Peccei-Quinn solution,
solves the CP problem and has cosmological implications.
In this scenario $\theta$ is therefore made local, where nonperturbative QCD
effects produce a mass for the axion and fix the background value
of $\theta_{QCD}$ to be zero.

The interaction of an axion field to electromagnetic fields
was initially investigated several years ago, after it was shown \cite{Witten}
that in a $\theta$ vacuum a t'Hooft-Polyakov magnetic monopole acquires
an electric charge proportional to $\theta$ and to the magnetic charge
$g$ of the monopole,
\beq
q_{\theta}=C\theta g
\eeq
with C a constant.
In particular the dynamics of monopoles traversing axionic domain walls
has since been elucidated \cite{Sikivie}, \cite{Huang}.

In its full generality, the interaction of the axion with the electromagnetic
field is assumed to be given, at low energies, by the action density
\cite{Sikivie}
\beq
L=-{1\over 4}F^2 +{1\over 2}\dmu\theta\partial^{\mu}\theta
-{{m^2 v^2}\over {N^2}}[1 - cos(Na/v)]
+\alpha F\tilde{F}
\eeq
where $\tilde{F}$ denotes the dual of the field strength and
\beq
\alpha = {e^2\over {32\pi^2 sin^2{\theta^{0}}_{w}}}
\left(\theta_{qcd} +{{ T_{\theta}}\over 2\pi}{{N\theta(x)}\over v}
\right)
\eeq
where $\theta(x)$ is the axion field, m is the axion mass,
N is the number of axion vacua and $v$ is the vacuum expectation value
 that breaks the
Peccei-Quinn symmetry.

${\theta^{0}}_{w}$ is the electroweak angle at the grand unification mass scale
$(GUM)$ and $\theta_{qcd}$, the constant  is taken at  GUM.

$T_{\theta} =2\pi$ for QCD,
is the period of $\theta$.

We have assumed that there is grand unification of the strong and
electroweak interactions \cite{Georgi} and we have used
$$g_s= e^2/sin^2({\theta^{0}}_{w}) \,\, $$
at GUM,
where $g_s$ is the QCD coupling constant at the same scale.

In this letter we investigate the propagation of high
frequency electromagnetic  modes
in the presence of an axion. We use the
light cone expansion of the propagator for the electromagnetic field,
and we calculate the local part of the one-loop effective action of the axion
field.
In the case of a slowly varying $\theta$ field, representing a massive
axion, we show that the coefficient of the leading singularity
 of the asymptotic expansion
can be explicitely determined in full
nonlocal form.
This last result allows us to define a causal gauge field
propagator in the
presence of a massive axion, a result which cannot be obtained
from the dispersion relation
because it contains
 spurious poles.
As noticed previously in the literature \cite{Jackiw}, this particular
background configuration of the axion field (the massive axion) makes the
propagation of
the
low frequency em modes unstable and a tachionic pole arises.
The investigation of ref.\cite {Jackiw} was carried out in the particular case
of a coupled
Maxwell-Chern-Simons theory, a system which shares the same properties of the
massive
axion model.

We show that, for any  background axion field,
and in particular for a massive axion model,
the em propagator behaves well at short distances.
Our results are  based on a direct application of the Hadamard
theorem \cite{Hadamard} for equations of propagation which are diagonal in the
highest derivatives.
The short distance behaviour of the model is in fact controlled by the
highest derivatives of the gaussian operator (which is diagonal),
while the instability is generated by the presence of lower
derivatives (coupled to an $\epsilon$ tensor) in the equations of
motion.

\newpage
The partition function of the model can be written in the simplified form
\beq
\label{one}
Z=\int [d\theta] [d A_{\mu}] e^{i \int {\cal{L}+ J\theta +J'_\mu A^\mu} d^{4}x}
\eeq
\beq
\label{two}
{\cal{L}}= -\quarter F^{2} + \half \dmu \theta \partial^{\mu} \theta +
{c\over 4} \theta (x) F \tilde{F}\eeq
$F$ denotes the electromagnetic field, $\tilde{F} = {1\over 2}\epsilon F$
denotes its dual.

We will work in the Lorentz gauge.  The evaluation
of the effective action for the $\theta$-field, functionally integrating
out the gauge fields is given by
\beq
\label{three}
{\cal{L}}_{eff} = \half \dmu \theta \partial^{\mu} \theta +\frac{i}{2} tr
\ln(\hat{P})
\eeq
where
\beq
\label{four}
\hat{P} = g^{\nu\beta} \Box  -c \dmu \theta \epsilon^{\mu \nu \alpha \beta}
\partial_{\alpha}
\eeq
is the relevant operator induced by the gaussian approximation.

The equations of motion for the gauge fields are:
\beq
\label{five}
\dmu F^{\mu \nu} =  c \dmu \theta \epsilon^{\mu \nu \alpha \beta}
\partial_{\alpha} A_{\beta}
\eeq
where
\beq
\label{six}
F_{\mu \nu} = \dmu A_{\nu} - \dnu A_{\mu}
\eeq

In the Lorentz gauge we find
\beq
\label{seven}
\Box A_{\nu} - k_{\nu \alpha \beta} \partial^{\alpha} A^{\beta}=0
\eeq
where we have defined
\beq
\label{eight}
k_{\nu \alpha \beta} =  c \dmu \theta \epsilon^{\mu \nu \alpha \beta}
\eeq

The Hadamard  expansion \cite{Hadamard} for the propagator of the gauge fields
in the
axionic background can be set directly by
introducing the Green's function $\hat{\Delta}$ for eq.(11),
since the gaussian operator is diagonal in the highest derivatives
\beq
\label{nine}
\Box \hat{\Delta}_{\nu \rho} - k_{\nu \alpha \beta} \partial^{\alpha}
\hat{{\Delta}^{\beta}}_{\rho} = \delta_{\nu \rho} \delta^{4}(z)
\eeq
where $z=x-y$.
The support of the $\Delta$ distribution is therefore inside the
light-cone and causality is respected.

The standard ansatz is therefore
\beqa
\label{ten}
\hat{\Delta}_{\nu \rho}(x,y) & = & G^{(0)}_{\nu \rho}(x,y) D_{F}  \nonumber \\
&   & \mbox{} - \frac{i}{16 \pi^{2} }
\ln({z^{2}\over {\mu}^2}-i 0) \sum_{n=0}^{\infty}
(\frac{z^2}{4})^{n}\frac{1}{n!} G_{\nu \rho}^{(n+1)}(x,y)
\eeqa
where
\beq
\label{eleven}
D_{F} = \frac{1}{4 \pi^{2} i (z^{2}-i 0)}
\eeq
is the Feynman free propagator in coordinate space while $\mu$ is a
mass parameter introduced in order to keep dimensionless the argument
of the logarithm in the expansion.

The recursion relations for the coefficients of the expansion are
 easily obtained.

By equating to zero the independent singularities of eq. (13) applied to the
 expansion (14) we get
a leading singularity equation for $z^{-4}$
\beqa
\label{thirteen}
2 z^{\mu} \dmu G^{(0)}_{\nu \rho}(x,y) - z^{\alpha} k_{\nu\alpha\beta}
G^{(0)}_{\beta\rho}(x,y) & =  & 0
\eeqa
a $D_{F}$ (i.e.$z^{-2}$) singularity equation:
\beqa
\label{fourteen}
\Box G^{(0)}_{\nu\rho} + G^{(1)}_{\nu\rho}+z^{\mu}\dmu G^{(1)}_{\nu\rho} -
k_{\nu\alpha\beta}\partial^{\alpha}G^{(0)}_{\beta\rho} &   &\nonumber \\
-{z^\alpha \over 2} k_{\nu\alpha\beta}{{G^{(1)}}^{\beta}}_{\rho} & = & 0
\eeqa
and a log-equation (for $z^{2n}ln({z^2\over {\mu}^2})$)

\beqa
\label{fifteen}
z^{\mu} \dmu G^{n+2}_{\nu\rho}+(n+2)G^{(n+1)}_{\nu\rho} +
\Box G^{(n+1)}_{\nu\rho} &   & \nonumber \\
\mbox{} - \frac{z^{\alpha}}{2} k_{\nu\alpha\beta}
G^{(n+2)\beta}_{\rho} -
k_{\nu\alpha\beta} \partial^{\alpha} G^{(n+1)}_{\beta\rho} & = & 0
\eeqa

Defining $G^{(-1)}= 0 $, $n=-1,0,1,...$, we summarize all the equations in the
form:
\beqa
\label{sixteen}
z^{\mu} \dmu G^{(n+1)}_{\nu\rho} + (n+1)G^{(n+1)}_{\nu\rho} + \Box
G^{(n)}_{\nu\rho}  &   & \nonumber \\
\mbox{} - \frac{z^{\alpha}}{2} k_{\nu\alpha\beta}
G^{(n+1)}_{\beta\rho} - k_{\nu\alpha\beta}\partial^{\alpha}G^{(n) \beta}_{\rho}
& = & 0
\eeqa

At this level the axion is a background field and we see, for instance
from eq. (16), that it affects the propagation of the
leading singularity in a non trivial way  (through
$k_{\nu\alpha\beta}$).

For a constant  $\theta$-angle the equation for the propagator of the
leading singularity is trivially given by

\beq
\label{seventeen}
z^{\mu}\dmu G^{(0)}_{\nu\rho} = 0
\eeq
valid on the light cone surface, with the initial condition
\beq
G^{(0)}_{\nu\rho} (x,x)=\delta_{\nu\rho}
\eeq

The solution is trivially given as the Kronecker delta, and the strength
therefore is diagonal over the entire characteristic surface.
 In this simplified case the equations of motion for the gauge fields are given
by
\beq
\label{eighteen}
\Box \Box^{-1} = \delta_{\nu \rho} \delta^{4}(z)
\eeq
where as usual,
\beq
\label{nineteen}
\Box^{-1} = \frac{G^{(0)}_{\nu\rho}}{({z^{2}-i 0})4{\pi}^2 i}
 = \frac{\delta_{\nu\rho}}{({z^{2}-i 0})4{\pi}^2 i}
\eeq

At this point we focus our attention
on eq (16), which describes the leading behaviour of the dynamics of the
electromagnetic field in the presence of a local vacuum angle.

We will preliminarly show that, in the case of a linearly growing
axion field, in the timelike case, some components of the leading
singularity tensor
 ${G^{(0)}}_{\nu\rho}$ are constant on the light cone surface.
The  reason appears to be quite simple and is valid both in the abelian
and in the nonabelian case.
In fact, by contraction of both terms of (16) with $\partial_{\nu}\theta$
and using a symmetry argument (the antisymmetry of k defined in eq (12) )
we get the equation
\beq
\partial_\nu \theta z^{\mu}\dmu {G^{(0)}}_{\nu\rho}(x(s),0)=0
\eeq
If we introduce  the parametrization
\beq
x^{\mu}(s)=s x^{\mu}, 0<s<1; \,\,y=0
\eeq

for a straight line inside the light cone surface
we can rewrite eq (24) in the form
\beq
\partial_{\nu}\theta(x){d\over ds}{G^{(0)}}_{\nu\rho}(x(s),0)=0
\eeq
Repeating the same procedure a second time one gets the equation
\beq
x^{\nu}{d\over ds}G^{(0)}_{\nu\rho}(x(s),0)=0
\eeq

It is simple to show that
if the variation of the $\theta$ field is linear in a  timelike direction,
 a frame
can be found in which it has only a time variation, say
\beq
\dmu\theta =(a_0,0,0,0)
\eeq
and in particular we get for the timelike components
\beq
{d \over ds}G^{(0)}_{0\rho} =0
\eeq
{}From the initial condition eq (21) therefore we get
\beq
{G^{(0)}}_{0 i}=0 \,\, \,\,\,{G^{(0)}}_{0 0}=1
\eeq

As we emphasized previously, this result remains true even in the more
realistic non abelian case, for a linearly growing (timelike) axion field,
and is a simple consequence of the antisymmetry involved in the
coupling. A simple and physical way to look at this result
is to view it as an anisotropy effect induced on the propagation of the gauge
fields by the pseudo tensor coupling.

We will show now that  the other components of the leading singularity
tensor can also be determined by following an approach analogous to the one
we used above, in the case in which condition (28) is enforced.

The equation for the leading singularity (eq. 16) can be cast in the form

\beq
{d\over ds}{G^{(0)}}_{ij}(x(s),0)
-{c a_0\over 2}x^k\epsilon^{0ikl}{G^{(0)}}_{lj}(x(s),0)=0 \\
\,\,\,\,\,\,i,j,k,l=1,2,3
\eeq

For a given timelike vector $x^\mu$, the previous equation has
two first integrals.
This can be easily seen through an elementary analogy with the
motion of a classical charged particle under a Lorentz force
in a constant magnetic field.
We use this analogy to solve it.
Define
\beq
G^{(0)}_{ij}(x(s),0)=
\hat{v_j}; \,\,\,\,{\omega}^k={a_0\over 2}x^k
\eeq
where $\hat{\omega}$ is constant at fixed $x^\mu$
and rewrite it as
\beq
{d\over ds}\hat{v_j}=\hat{\omega}\wedge\hat{v_j}
\eeq
and two first integrals of motion along the s-line are

\beq\hat{\omega}.\hat{v_j}={a_0\over 2}x^k G^{(0)}_{kj}
(x(s),0)=d_j
\eeq

\beq \hat{{v^2}_j}=\sum_{i} G^{(0)}_{ij}(x(s),0)G^{(0)}_{ij}
(x(s),0)=l_j
\eeq

Iterating eq.(33) we get
\beq
\left({d^2\over ds^2}+{\omega}^2 \right)\hat{v_j}=
\hat{\omega}d_j
\eeq

By usingff the initial condition eq (21) one easily gets
\beq
d_j={c a_0\over 2}x^j \,\,\,\, l_j=1,\,\,j=1,2,3
\eeq
and finally the solution
\beq
G^{(0)}_{ij}(x(s),0)=\left(\delta_{ij}-{ x^i x^j\over r^2} \right)
cos({c a_0\over 2}rs) +\epsilon_{ikj}{x^k\over r}
sin({c a_0\over 2}rs) + {x^i x^j\over r^2}
\eeq
This expression describes the behavior of the gauge field correlator
in its non local form, asympotically around the light-cone.

Our results can be summarized in the following expressions
for the Feynman propagators of electromagnetic fields in the
linearly growing timelike $\theta$ vacuum

$${<A_{i}(x)A_{j}(0)>}_{\theta}=$$
\beq
\left[ (\delta_{ij}- {x^{ij}\over r^2})cos({{a_0r}\over 2})
+\epsilon^{ikj}{x^k\over r}sin({{a_0r}\over 2})+ {x^ix^j\over r^2}\right]
{1\over {4\pi^2 i(x^2-i0)}} +\,\, log.\,terms
\eeq

\beq
{<A_{0}(x)A_{i}(0)>}_{\theta}
={{\delta_{0i}}\over {4\pi^2i(x^2-i0)}} +
G^{(1)}_{0i}(x,0)log(x^2-i0)
\eeq
modulo additional logarithmic corrections.
This model provides an additional example of a solution
for the Green function in an external field in full nonlocal form.
The outline of this approach, as is well known, dates back
to Schwinger \cite{Schwinger} and it dealt with the Dirac propagator in an
abelian gauge
field. Causality problems have been investigated by Velo and
Zwanziger \cite{Zwanziger} in this
framework successfully, and calculations of anomalies
are also possible.
The massive axion model suffers, however, of an instability at low
frequencies,
in other words the retarded em propagator developes a tachionic
behaviour in its lower modes.
This aspect has been considered in very detail in ref. \cite{Jackiw}
in which the 3+1 dimensional Maxwell-Chern-Simons (MCS) system is discussed.
The equations of motion for the propagation of electromagnetic fields
in the massive axion background are equivalent, in fact, to the MCS
model. An expression of the retarded propagator was also given in it and it
was shown that em frequencies smaller than $a_0$, the time component
of the derivative of the axion field, becomes tachionic \cite{Jackiw}.
We refer to that paper for further details.
In the MCS case the instability is an inevitable effect, since no
background
field is involved from the beginning, while in the massive axion model
is still unclear whether by taking into account in a better way the
axion dynamics the instability can be removed.

In its current formulation, in the case of
either a spacelike or timelike variation
of the local vacuum, the  propagation of the gauge fields appears
to be consistent
in the ultraviolet regime  and, therefore, the singular part of the
effective action for the axion field, obtained by integrating out the
gauge fields in the partition function, can be computed straightforwardly.


In the following we are going to show that the structure of the
singularities of the propagator of the gauge field in an
arbitrary $\theta$ background has a very simple feature.

\section{The effective action}
In the general case of a variable vacuum angle we can easily calculate
the singular (local) part of the effective action for the
$\theta$ field integrating out the gauge field.
Let's define $[G^{(n)}_{\mu \nu}]$ to be the coincidence limit of the
coefficients introduced before.
\beq
\label{twenty}
[G^{n}_{\mu \nu}] \equiv G^{(n)}_{\mu\nu}(x,x)
\eeq
The same coefficients also appear
 in the divergent part of the one loop effective action
\beq
\label{twentyone}
\Gamma^{(1)}(\theta) = \frac{i}{2} tr \ln(\hat{P})
\eeq

Let's take a variation of (42) with respect to the background field
$\theta $

\beq
\delta \Gamma^{(1)}[\theta] = \frac{i}{2} tr \delta P P^{-1} =
\frac{i}{2} <\delta\hat{P}_{\mu\nu}\hat{\Delta}^{\nu\mu}(x,y)\delta^{4}(z)>
\eeq
where the angular bracket denotes integration over x,y.

We find after some algebra
\beq
\label{twentythree}
[\delta \hat{P}_{\nu\beta} \hat{\Delta}^{\beta\nu}]=
- c <\dmu \delta\theta \epsilon^{\mu\rho\alpha\nu} \partial_{\alpha}
\Delta_{\nu \rho}(x,y)\delta^{4}(z)>
\eeq

Ultraviolet divergences in the effective action come from the singular
behaviour of the propagator at short distances.
 We need to regularize expressions such as $[\ln(z^{2}-i 0)]$, $[D_{F}]$ and
$[\dmu D_F] $. This  can be done in many ways.
The simplest one consists in introducing a short distance cutoff
$\Lambda$. All the singularities are regulated at the intermediate
stage, while
we first compute the coincidence limits of the various coefficients.
The cutoff is removed by taking the limit $\Lambda \rightarrow \infty$
at the end.

A simple calculation shows that the relevant contribution to (44)
is given by
\beq
\label{twentyfour}
[\partial_{\alpha}\hat{\Delta}{\nu\rho}]= -\frac{i}{16 \pi^{2}}
[\ln(z^{2}-i 0)][\partial_{\alpha}G^{(1)}_{\nu\rho}] +
[\partial_{\alpha}G^{(0)}_{\nu\rho}][D_{F}]+
[{G^{(0)}}_{\nu\rho}] [\partial_{\alpha} D_F]
\eeq

For notational simplicity we omitted to write down  in eq. (45) explicitely
 the
cutoff dependence of the singular coincidence limits.

It can be seen immediately that, in this regularization, the contribution from
the $ [\partial D_F]$
 singularity is zero.

By differentiating the singularities equations we find
\beq
\label{twentyfive}
[\partial_{m} G^{(0)}_{\nu\rho}] = \half k_{\nu m \rho}
\eeq
\beqa
\label{twentysix}
[\partial_{m} \partial_{n} G^{(0)}_{\nu\rho}] & = & \quarter
\partial_{n} k_{\nu m \rho} + \quarter \partial_{m} k_{\nu n \rho}
\nonumber \\ &   & {1\over 8}{k_{\nu m}}^{\beta}k_{\beta n \rho}+
{1\over 8}{{k_{\nu n}}}^{\beta}k_{\beta m \rho}
\eeqa

The explicit expression for the counterterm is
\beq
\label{thirty}
<[\delta\hat{P}_{\tau\nu}\hat{\Delta}^{\nu\tau}]>=\int d^4x
\delta\dmu\theta
\epsilon^{\mu\tau\alpha\nu}\left(\frac{-c}
{32 \pi^{2} i}[\ln(z^{2}-i 0)][\partial_{\alpha}G^{(1)}_{\nu\tau}]
+ [\partial_{\alpha} {G^{(0)}}_{\nu\rho}] [D_F] \right)
\eeq

A straightforward  but lenghty calculation gives

\beq
\Gamma^{(1)}(\theta)={i\over 2} \int d^4 x \left( {-c\over 32\pi^2 i}
[ln z^2] ({1\over 4} (Q^2)^2 c^2 -{1\over
2}\dmu\theta\Box \partial^{\mu}
\theta)+{1\over 4\pi^2 i}[{1\over z^2}] {3 c\over 2} Q^2 \right)
\eeq

where we've defined
\beq
Q^2=\dmu \theta\partial^{\mu}\theta
\eeq
Introducing a pole and a logarithmic renormalization costants $ Z_1$
and $Z_2$, we can express the counterterm lagrangean in the form

\beq
\label{thirtyfive}
\nabla {\cal{L}}^{count.} =c Z_1 Q^2
+ c Z_2 \left(c^2 (Q^2)^2 - 2 \dmu\theta\Box\partial^\mu\theta \right)
\eeq
\section{\bf Conclusions}
 We have analyzed the dynamics of a coupled axion-gauge field model
in the low energy limit and we've investigated in detail the problem of the
propagation of high frequency em modes in an axion background.
In particular we have seen that, in the case of a timelike varying axion
background, the leading singularity of the Hadamard
expansion of the gauge fields correlator can be determined in
non local form.
By applying Hadamard's theorem on the structure of the
singularities of diagonal hyperbolic operators, we have shown that the
instability of the Maxwell-axion system is
confined to the propagation of low em frequencies, reobtaining a result
first considered in ref. \cite {Jackiw}. Our results are valid for the
propagation of em modes in any axionic
background. We have also shown that the local part of the effective action of
the axion can be
computed
in a simple form, a result which gives to the effective  model physical
consistency.

\vspace{.5cm}

\newpage

\centerline{\bf Acknowledgments}
\vspace{.5cm}

I acknowledge Profs. A.S. Goldhaber, H. Yamagishi, B. W. Lindquist
 J.J.M. Verbaarschot and G. Sterman. I warmly thank Prof. D. Zwanziger
and Dr. F. Bastianelli for  clarifying discussions and for helpful
suggestions. I finally thank the Theory Group at Univ. of Lecce, Italy
for their kind hospitality, Susan Marie Trapasso and Rinat Kedem for
their generous help.

\end{document}